\documentclass[aps,pre,noshowkeys,
nosuperscriptaddress,
longbibliography,
reprint,
twocolumn, 
nofootinbib,
]{revtex4-1}

\usepackage{amsmath}
\usepackage{subcaption}
\usepackage{graphicx}
\usepackage{tikz}
\usepackage{wasysym}

\usepackage[colorlinks=true]{hyperref}
\usepackage{cleveref}

\begin{document}

\title{Comment on `Phase transition in a network model of social balance with Glauber dynamics'}

\author{Krzysztof Malarz}
\thanks{\includegraphics[width=10pt]{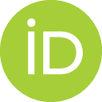}~\href{https://orcid.org/0000-0001-9980-0363}{0000-0001-9980-0363}}
\email{malarz@agh.edu.pl}
\author{Krzysztof Ku{\l}akowski}
\thanks{\includegraphics[width=10pt]{ORCID.png}~\href{https://orcid.org/0000-0003-1168-7883}{0000-0003-1168-7883}}
\email{kulakowski@fis.agh.edu.pl}
\affiliation{AGH University of Science and Technology,
Faculty of Physics and Applied Computer Science,
al. Mickiewicza 30, 30-059 Krak\'ow, Poland}

\date{\today}

\begin{abstract}
In a recent work [R. Shojaei {\it et al}, Physical Review E {\bf 100}, 022303 (2019)] the Authors calculate numerically the critical temperature $T_c$ of the balanced-imbalanced phase transition in a fully connected graph.
According to their findings, $T_c$ decreases with the number of nodes $N$. 
Here we calculate the same critical temperature using the heat-bath algorithm. 
We show that $T_c$ increases with $N$ as $N^{\gamma}$, with $\gamma$ close to 0.5 or 1.0. 
This value depends on the initial fraction of positive bonds.
\end{abstract}

\maketitle

The concept of structural balance (Heider balance) is well established in social psychology, and it has 
counterparts in computational science, in particular in simulations on networks \cite{antal}. 
Sites $i=1,\cdots,N$ in a network represent actors, and bonds $x_{ij}=x_{ji}$ represent relations between them.
For friendly relations $x_{ij}=+1$, and for hostile ones $x_{ij}=-1$.
In each balanced state $x_{ij}x_{jk}x_{ki}=+1$ for each triad $ijk$.
Departures from the balanced state are usually calculated {\it via} the mean value of the product $x_{ij}x_{jk}x_{ki}$.
The evolution should drive the network towards balance, and various algorithms have been designed with this purpose \cite{antal,2005.11402}.

\begin{figure*}
\centering
\begin{subfigure}[b]{.33\textwidth}
\caption{}
\includegraphics[width=.99\textwidth]{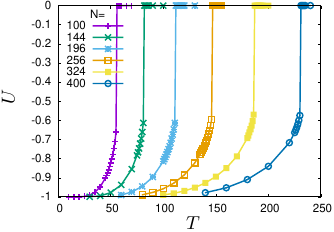}
\end{subfigure}
\begin{subfigure}[b]{.32\textwidth}
\caption{}
\includegraphics[width=.99\textwidth]{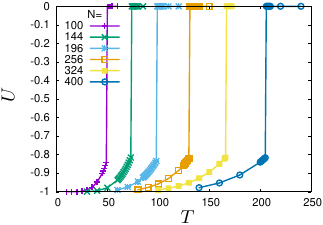}
\end{subfigure}
\begin{subfigure}[b]{.32\textwidth}
\caption{}
\includegraphics[width=.99\textwidth]{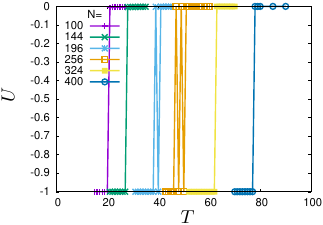}
\end{subfigure}
\begin{subfigure}[b]{.32\textwidth}
\caption{}
\includegraphics[width=.99\textwidth]{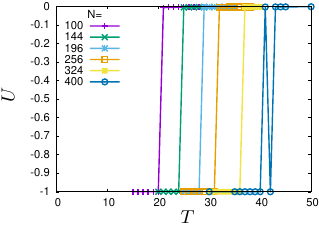}
\end{subfigure}
\begin{subfigure}[b]{.32\textwidth}
\caption{}
\includegraphics[width=.99\textwidth]{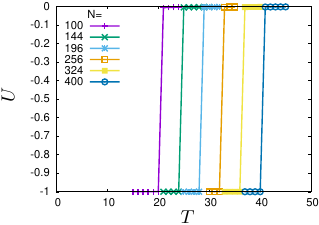}
\end{subfigure}
\begin{subfigure}[b]{.32\textwidth}
\caption{}
\includegraphics[width=.99\textwidth]{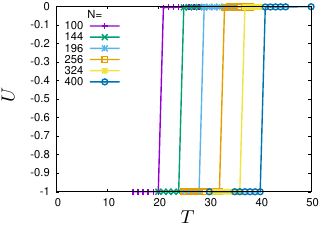}
\end{subfigure}
\begin{subfigure}[b]{.32\textwidth}
\caption{}
\includegraphics[width=.99\textwidth]{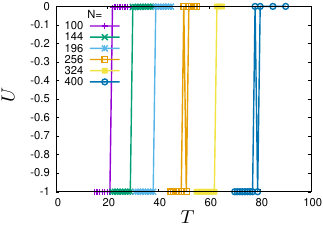}
\end{subfigure}
\begin{subfigure}[b]{.32\textwidth}
\caption{}
\includegraphics[width=.99\textwidth]{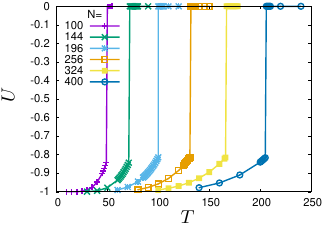}
\end{subfigure}
\begin{subfigure}[b]{.33\textwidth}
\caption{\label{fig:UvsTrho0100}}
\includegraphics[width=.99\textwidth]{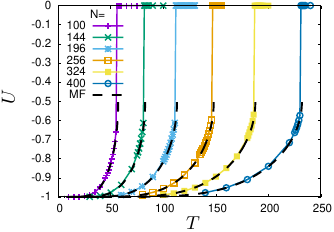}
\end{subfigure}
\caption{\label{fig:UvsT}(Color online). System energy $U=-\sum x_{ij}x_{jk}x_{ki}/\binom{N}{3}$ dependence on temperature $T$ for various system size $N$ and various initial concentration $\rho_0$ of positive bonds: (a) $\rho_0=0$, (b) $\rho_0=0.2$,  (c) $\rho_0=0.4$, (d) $\rho_0=0.45$, (e) $\rho_0=0.5$, (f) $\rho_0=0.55$, (g) $\rho_0=0.6$, (h) $\rho_0=0.8$, (i) $\rho_0=1$. In \Cref{fig:UvsTrho0100} the dashed black lines show mean-field approximation results for $\rho_0=1$ based on Ref.~\onlinecite{1911.13048}. For each value of $N$, the mean-field results coincide with those from our simulations.}
\end{figure*}

\begin{figure*}
\begin{subfigure}[b]{.59\textwidth}
	\caption{\label{fig:TCvsNq_a}}
\includegraphics[width=.99\textwidth]{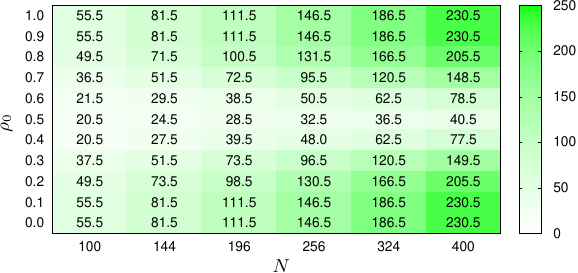}
\end{subfigure}
\begin{subfigure}[b]{.39\textwidth}
	\caption{\label{fig:TCvsNq_b}}
\includegraphics[width=.99\textwidth]{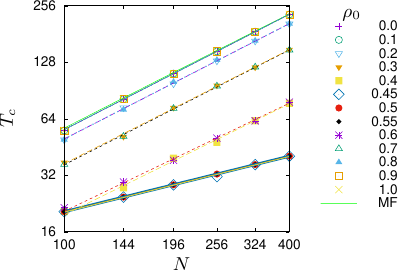}
\end{subfigure}
\caption{\label{fig:TCvsNq}(Color online). Critical temperature $T_c$ for various initial concentration of positive bonds and various system size $N$. (a) $T_c(N;\rho_0)$, (b) $T_c(N)$. In \Cref{fig:TCvsNq_b} the solid green line shows mean-field approximation results for $T_c(N;\rho_0=1)$ \cite{1911.13048}.}
\end{figure*}

Recently the problem has been treated with methods of equilibrium statistical mechanics, with the mean value of $x_{ij}x_{jk}x_{ki}$ as energy $U$ \cite{PhysRevE.99.062302,PhysRevE.100.022303,1911.13048}.
In particular, the critical temperature $T_c$ has been calculated numerically for the balanced-imbalanced phase transition. 
For $T<T_c$ the energy $U$ is close to $-1$, while above $T_c$ its value is near zero. 
Our comment is on the dependence of $T_c$ on the system size $N$. 
In Ref.~\onlinecite{PhysRevE.100.022303}, this dependence is shown on Figure 6a; there, $U$ is presented as function of thermodynamic beta $\beta=1/T$ for different values of $N$.
When $N$ increases, the jump of $U$ from zero down to $-1$ is shown for larger values of $\beta$; this means that $T_c$
{\em decreases} with $N$. 

The results of Ref.~\onlinecite{PhysRevE.100.022303} are obtained with the Glauber dynamics. 
Below we present results of our simulations of $T_c(N)$ with using the heat-bath algorithm. The time evolution of a link $x_{ij}(t)$ is given by the rule 
\begin{subequations}
\label{eq:evol}
\begin{equation}
\label{eq:evol_x}
x_{ij}(t+1)=
        \begin{cases}
	+1 & \text{ with probability }p_{i,j}(t),\\
	-1 & \text{ with probability }[1-p_{i,j}(t)],
        \end{cases}
\end{equation}
where $p_{i,j}(t)$ is given as
\begin{equation}
\label{eq:evol_p}
    p_{i,j}(t)=\frac{\exp[\xi_{i,j}(t)/T]}{\exp[\xi_{i,j}(t)/T]+\exp[-\xi_{i,j}(t)/T]} 
\end{equation}
and 
\begin{equation}
\label{eq:evol_xi}
	\xi_{i,j}(t)=\sum_{k\ne i,j} x_{ik}(t) x_{kj}(t).
\end{equation}
\end{subequations}
\Cref{eq:evol} is applied synchronously to all edges.

\begin{table*}[htbp]
\caption{\label{tab:gamma}Exponent $\gamma$ and its uncertainty $u(\gamma)$ as dependent on initial density $\rho_0$ of positive bonds.}
\begin{ruledtabular}
\begin{tabular}{r rrrrrrrrrrrrr}
$\rho_0$   &    0.0&    0.1&    0.2&   0.3&   0.4&  0.45&     0.5&    0.55&   0.6&   0.7&    0.8&    0.9&    1.0\\ \hline
$\gamma$   & 1.0210& 1.0210& 1.0208& 1.010& 0.983& 0.487& 0.49155& 0.48672& 0.958& 1.017& 1.0208& 1.0210& 1.0210\\
$u(\gamma)$& 0.0032& 0.0032& 0.0069& 0.015& 0.029& 0.014& 0.00044& 0.00096& 0.020& 0.010& 0.0088& 0.0032& 0.0032\\
\end{tabular}
\end{ruledtabular}
\end{table*}

Below we show that the critical temperature $T_c$ depends on the density $\rho_0$ of positive bonds at $t=0$ as
\begin{subequations}
\begin{equation}
T_c(N;\rho_0)\propto N^{\gamma(\rho_0)}
\end{equation}
where
\begin{equation}
\gamma(\rho_0)\approx\begin{cases}
\frac{1}{2} & \text{ for } \left|\rho_0-\frac{1}{2}\right|<0.1,\\
1           & \text{ for } \left|\rho_0-\frac{1}{2}\right|>0.1.
\end{cases}
\end{equation}
\end{subequations}

\begin{figure*}
\begin{subfigure}[b]{.48\textwidth}
\caption{\label{fig:rho_vs_t_T20}}
\includegraphics[width=.99\columnwidth]{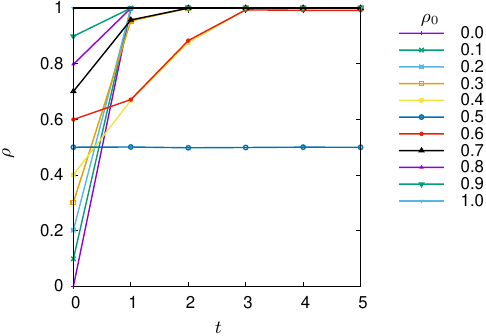}
\end{subfigure}
\hfill 
\begin{subfigure}[b]{.48\textwidth}
\caption{\label{fig:rho_vs_t_T120}}
\includegraphics[width=.99\columnwidth]{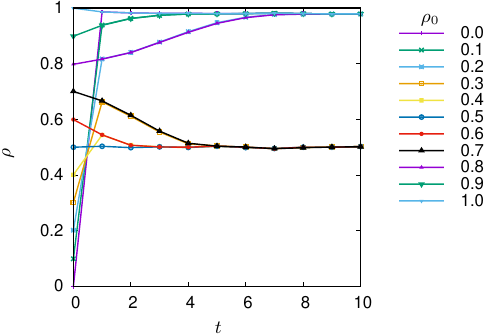}
\end{subfigure}
\caption{\label{fig:rho_vs_t}Time evolution of density $\rho(t)$ of positive links for $N=256$, (a) $T=20$, (b) $T=120$.}
\end{figure*}

Summarizing our numerical results, the data shown in \Cref{fig:UvsT,fig:TCvsNq} indicate that the critical temperature $T_c$ {\em increases} with the system size $N$. Namely, $T_c \propto N^{\gamma}$, where $\gamma>0$. The exponent $\gamma$ depends on the initial fraction $\rho_0$ of positive links. Within the numerical accuracy, the results on $T_c$ are symmetric with respect to $\rho_0=0.5$, {\it i.e.} $\gamma(\rho_0)=\gamma(1-\rho_0)$---see \Cref{tab:gamma}. This symmetry is due to the fact that both two positive and two negative bonds $x_{jk}$ and $x_{ki}$ contribute to a positive value of $x_{ij}$ in the next time step.  Once there is an excess of bonds of a given sign (plus or minus), in the next step  positive bonds prevail. This is shown in \Cref{fig:rho_vs_t}, where the density $\rho$ of positive bonds tends to 1, if only its initial value is not too close to 0.5. (The latter case is discussed later in the text.) Further, if the mean value of $x_{ij}$ is different from zero, adding a $n$-th node enhances the effective field $\xi_{ij}$ by $x_{in}x_{nj}$, which is more often positive, and therefore the critical temperature increases linearly with the system size.

The case when the mean value of $x_{ij}$ is close to zero ($\rho_0\approx 0.5$) is different. Our numerical results show that $\rho= 0.5$ is a stable fixed point of the time evolution. This is because near this point the product $x_{ik}x_{kj}$ is positive or negative with equal probabilities, the values of $\xi_{ij}$ oscillate around zero, and the mean value of $p_{ij}(t)$ is 0.5; hence $x_{ij}=\pm 1$ with the same probabilities also in the next time step.  Then the absolute value of the mean local field $\xi_{ij}$ should be evaluated from the standard deviation of its distribution. The latter increases with $N$ as $\sqrt{N}$.

These arguments find support also when we compare the results of \Cref{fig:TCvsNq_a} with the data in \Cref{fig:rho_vs_t_T20}. 
The temperature $T=20$ is lower than the critical temperatures for all values of $\rho_0$. 
As we argued above, the case $\rho=0.5$ is a fixed point, yet except for this value, all curves $\rho(t)$ tend to 1.0. 
On the contrary, $T=120$ is higher than $T_c$ for $\rho_0$ between 0.25 and 0.75, and $\rho(t)$ tend to 0.5 precisely for these cases (\Cref{fig:rho_vs_t_T120}). 
This is an indication, that for $T>T_c$ the thermal noise restores the symmetry of the distribution of $x_{ij}$ around zero.

The state with $\rho=0.5$ is absorbing, what can be demonstrated in the following thought experiment. 
We start the system at low temperature (below $T_c$) and $\rho_0=0.8$. 
As shown in \Cref{fig:rho_vs_t_T20}, the density $\rho$ increases to almost one. 
Then we heat the system above $T_c$, which is high (\Cref{fig:TCvsNq}). 
In these conditions $\rho$ tends to 0.5. 
To reach the balanced state, we have to cool the system down below the critical temperature, which is lower now. 
Yet further manipulations with temperature do not modify the density $\rho$, which remains equal to 0.5.

The increase of $T_c$ with $N$ is confirmed also in Ref.~\onlinecite{1911.13048} by simulations with the heat-bath algorithm for $\rho_0=1$, ({\it i.e.} in the vicinity of Heider's paradise; Table 1 there) and by the crude mean-field approximation yielding 
$T_c\approx(N-2)/1.71649$.

Main point of this comment is that the increase of $T_c$ with $N$ reported above is in contradiction with the result of Ref.~\onlinecite{PhysRevE.100.022303}. The origin of this contradiction is that the Authors of Ref.~\onlinecite{PhysRevE.100.022303} have used the energy per a triad. Roughly, they divided the total energy by $N^3$, as in their Eq. (1). The point is that to calculate the Monte Carlo probabilities they used the same rescaled energy (their Eq. (2)). Accordingly, their critical temperature is rescaled in the same way. This choice of scale is different from the appropriate Monte Carlo approach \cite[p. 8]{Newman1999}, where a non-normalized energy is used. We note that in our \Cref{eq:evol} the probabilities are calculated without the normalization. 

In our approach the system Hamiltonian \cite{Antal_2005,PhysRevE.99.062302} is
\begin{equation}
\label{eq:H-Heider}
\mathcal{H}=-\sum_{i,j>i,k>j} x_{ij}x_{ik}x_{jk},
\end{equation}
what corresponds to links updating scheme given in \Cref{eq:evol}.
Manipulation in additional factors in \Cref{eq:H-Heider} must lead to change of critical temperature $T_c$. 
Note, that in the classical Ising model on square lattice the Hamiltonian 
\begin{equation}
\label{eq:H-Ising}
\mathcal{H}=-J\sum_{i,j>i } s_is_j,
\end{equation}
where spin variables $s_i=\pm 1$, yields $T_c\approx 2.27$ but only when temperature is expressed in $J/k_B$ units, what is usually achieved by setting both, the coupling constant $J$ and the Boltzmann constant $k_B$ equal to unity. In other words, setting $J\ne 1$ redefine (critical) temperature by a factor of $J$, and the same effect should be observed here for manipulation with Hamiltonian \eqref{eq:H-Heider}.

The rescaling of energy used in Ref.~\onlinecite{PhysRevE.100.022303} might be convenient unless the thermal properties are calculated against the system size. As such, it has been also used  in literature \cite{PhysRevLett.103.198701}. However, if used to calculate $T_c(N)$, this rescaling changes the results.  When discussing, which form of energy is more appropriate, one should take into account the character of the simulated process. Here we discuss the social process of a removal of the structural imbalance, as described by Fritz Heider in 50’s. In our opinion, in this case the rescaling used in \cite{PhysRevE.100.022303} is incorrect.

The argument is as follows. Provided that a change of relation is to be decided by Alice towards Bob, how important is the number of other agents in the whole network? In a simplest case, there are at most two other agents, say Charlie and Denis. For $N=3$, what only matters for Alice is the product of actual relation between her and Charlie, multiplied by the relation among Charlie and Bob. For $N=4$, what does matter is also a product of relation between Alice and Denis, multiplied by the relation between Denis and Bob. Now, the issue is: should  the influence of Charlie and Denis be the sum of contributions of Charlie and Denis, or rather the average of these contributions? In other words, is the energy $U$ relevant for this social process additive, or not? Are two persons more influential than one? Our opinion that two agents are more influential is consistent with classical sociological experiments \cite{Asch1955} and with current sociophysical theories \cite{Sznajd-Sznajd-Sznajd}.

On the other hand, we do not claim that the updating scheme is irrelevant for the outcome of the Monte Carlo simulations. We agree that its influence could be a matter of a careful discussion in the context of particular aspects of the social process and its measurement. The importance of the updating has been demonstrated in social simulations \cite{PhysRevE.91.012108}. Moreover, the unifying Galam scheme \cite{cond-mat_0409484} for opinion dynamics predicts various critical temperatures $T_c$ if Metropolis or Glauber dynamics is applied \cite{Sousa2005}. Yet the choice of the updating scheme cannot change the character---increasing or decreasing---of the critical temperature dependence on the system size.

\begin{acknowledgments}
The authors are grateful to Pouya Manshour and Afshin Montakhab for helpful comments.
\end{acknowledgments}

%
\end{document}